\documentclass{emulateapj}

\usepackage{amsbsy}

\newcommand{\EQ}{\begin{equation}}
\newcommand{\EN}{\end{equation}}
\newcommand{\Eq}[1]{equation (\ref{#1})}

\newcommand{\Eqss}[2]{equations (\ref{#1})--(\ref{#2})}
\newcommand{\Sec}[1]{Sect.~\ref{#1}}
\newcommand{\Fig}[1]{Fig.~\ref{#1}}
\newcommand{\Tab}[1]{Table \ref{#1}}
\newcommand{\vc}[1] {{\boldsymbol #1}}
\newcommand{\dpa} {\partial}
\newcommand{\nab} {\vc{\nabla}}
\newcommand{\de} {{\rm d}}
\DeclareMathSymbol{\varOmega}{\mathord}{letters}{"0A}
\DeclareMathSymbol{\varSigma}{\mathord}{letters}{"06}

\shorttitle{Gravoturbulent Formation of Planetesimals}
\shortauthors{Johansen et al.}

\begin{document}


\title{Gravoturbulent Formation of Planetesimals}


\author{A. Johansen, H. Klahr and Th. Henning}
\affil{Max-Planck-Institut f\"ur Astronomie, K\"onigstuhl 17, 69117 Heidelberg,
Germany}




\begin{abstract}
  We explore the effect of magnetorotational turbulence on the dynamics and
  concentrations of boulders in local box simulations of a sub-Keplerian
  protoplanetary disc. The solids are treated as particles each with an
  independent space coordinate and velocity. We find that the turbulence has
  two effects on the solids. 1) Meter and decameter bodies are strongly
  concentrated, locally up to a factor 100 times the average dust density,
  whereas decimeter bodies only experience a moderate density increase. The
  concentrations are located in large scale radial gas density enhancements
  that arise from a combination of turbulence and shear.  2) For meter-sized
  boulders, the concentrations cause the
  average radial drift speed to be reduced by $40\%$. 
  We find that the densest clumps of solids
  are gravitationally unstable under physically reasonable values for the gas
  column density and for the dust-to-gas ratio due to sedimentation. We speculate that planetesimals can form in a 
  dust layer that is not in itself dense enough to undergo gravitational
  fragmentation, and that fragmentation happens in turbulent density
  fluctuations in this sublayer.
\end{abstract}




\section{INTRODUCTION}

Planets are believed to form from micrometer-sized dust
grains that grow by collisional sticking in protoplanetary gas discs (\citealt{Safronov1969}, see reviews
by \citealt{Lissauer1993} and \citealt{Beckwith+etal2000}). Once the bodies
reach a size of around one kilometer, the growth to Moon-sized protoplanets and
later real planets is achieved by gravitationally induced collisions
\citep{Thommes+etal2003}. Although
significant progress has been made in the understanding of the initial
conditions of grain growth \citep{Henning+etal2005}, we nevertheless do not yet
have a complete
picture of how the solids grow 27 orders of magnitude in mass to form
kilometer-sized planetesimals.

Growth by coagulation can take place when there is a relative speed
between the solids. Various physical effects induce relative speeds at
different grain size scales. This allows for a definition of distinct
steps in the growth from micrometer dust grains to meter-sized boulders in a
turbulent protoplanetary disc. Microscopic dust grains gain
their relative speed due to Brownian motion. 
This process forms relatively compact cluster-cluster
aggregates \citep{DominikTielens1997}. The speed of the Brownian motion falls
rapidly with increasing grain mass, and so the time-scale for building up
larger compact bodies this way becomes prohibitively large, compared to the
life-time of a protoplanetary disc.

When Brownian motion is no longer important, the relative speed is dominated by
the differential vertical settling in the disc. The vertical component of
the central star's gravity causes the gas to be stratified. Dust grains do
not feel the pressure gradient of the gas and thus continue to fall towards the
mid-plane with a velocity given by the balance between vertical gravity and the
drag force. Larger grains fall faster than smaller grains due to the
size-dependent coupling to the gas (actually bodies that are so massive that they
are starting to decouple from the gas will rather move
on inclined orbits relative to the disc, i.e.\ perform damped oscillations around the mid-plane). As they fall, they are thus able to sweep up smaller
grains in a process that is qualitatively similar to rainfall in
the Earth's atmosphere. Upon arrival at the mid-plane, the largest solids can
reach sizes of a few centimeters \citep{Safronov1969}. These bodies have grown as
compact particle-cluster aggregates with a high porosity.

Turbulent gas motions cause the sedimented
solids to diffuse away from the mid-plane
\citep{Cuzzi+etal1993,Dubrulle+etal1995}, where they can meet
and collide with a reservoir of microscopic grains. These tiny grains still
hover above the mid-plane because their sedimentation time-scale is so long
that turbulent diffusion can keep them well-mixed with the gas over a large
vertical extent. Turbulence also
plays a role for equal-sized macroscopic bodies by inducing a relative collision speed that
is much larger than the
Brownian motion contribution \citep{Voelk+etal1980,Weidenschilling1984}.

When estimating the outcome of an interaction between macroscopic bodies, the
issues of collision physics must be taken into account. For relative
speeds above a certain threshold, the bodies are likely to break up when they
collide rather than to stick \citep{Chokshi+etal1993,BlumWurm2000}. This is a
problem for macroscopic bodies where the sticking threshold is
a few meters per second.
Fragmentation caused by high-speed encounters continuously replenish the
reservoir of microscopic dust grains. These can then be swept up by the
boulders that are lucky enough to avoid critical encounters. However, the
sweeping up of smaller dust grains by a macroscopic body has its limitations when
the relative speed exceeds some 10 meter per second \citep{Wurm+etal2001}. At
larger relative velocities of up to a hundred meters per second, which are
likely to occur due to the high speed of larger bodies, the small grains will
erode the boulder. 

The time evolution of the size-distribution of solids can be calculated by
solving the coagulation equation numerically
\citep[e.g.][]{Wetherill1990,Weidenschilling1997,SuttnerYorke2001}. Recently,
\cite{DullemondDominik2005} performed numerical simulations of the coagulation
for realistic disc environments. Starting with micrometer-sized grains only,
they find that a narrow peak of 0.1--10 meter-sized boulders can form in
$10^4$--$10^5$ years, when fragmentation is ignored.
On the other hand, in a more realistic situation high speed impacts lead to
fragmentation. Here \cite{DullemondDominik2005} find that once the size distribution reaches the meter
regime, still around $75\%$ of the mass is maintained in microscopic bodies,
which are the fragments of larger bodies that have been destroyed in collisions.
This picture is given
some credit by the fact that microscopic dust grains are observed in
protoplanetary discs of millions of years of age, whereas the time-scale for
depleting grains of those sizes is only around 1,000 years
in the absence of fragmentation. 

Besides the problem of getting macroscopic bodies to stick, meter-sized boulders 
quickly drift radially inward toward the central star due
to their aerodynamic friction with the gas in a typical sub-Keplerian disc
\citep{Weidenschilling1977}. The drift time-scale can be as short as 100 years. To
avoid evaporation in the inner disc or in the central star, the bodies must
grow by least an order of magnitude in size (three orders of magnitude in mass)
in a time shorter than this!

A possibility to overcome the growth obstacles was suggested independently by
\cite{Safronov1969} and by \cite{GoldreichWard1973}.  The general idea is that
boulders sediment towards the mid-plane and form a particle sublayer that
undergoes a gravitational instability, forming the planetesimals in a
spontaneous event (\textit{gelation}) rather than by continuous growth
(\textit{coagulation}). The weakest point in this
model is that it requires a laminar disc in order to work. Even a tiny amount
of turbulence in the disc will prevent the boulders from an efficient
sedimentation towards the mid-plane, and the instability will never occur
\citep{WeidenschillingCuzzi1993}. Thus disc turbulence had always to be avoided
in order to allow for self-gravity assisted planetesimal formation. However,
even in a completely laminar disc, the settled dust induces a vertical shear in
the gas rotation profile
\citep{Weidenschilling1980,Nakagawa+etal1986}. This can be
unstable to a Kelvin-Helmholtz instability. The subsequent Kelvin-Helmholtz
turbulence puffs up the dust layer so that the densities needed for a
gravitational instability are usually not achieved, unless a dust-to-gas ratio
many times higher than the solar composition is adopted
\citep{YoudinShu2002}.

Nevertheless, solids can reach sizes of around one meter without the help of
self-gravity. In this size regime the gradual decoupling from the gas motion
enables the bodies to move independently from the gas. This can cause them to
be trapped in turbulent features of the gas flow. An important theoretical
discovery is that meter-sized boulders are concentrated in gaseous
anticyclonic vortices
\citep{BargeSommeria1995,Chavanis2000,Johansen+etal2004}. Inside such vortices
the dust density can locally be enhanced to values sufficient either for
enhanced coagulation or even for gravitational fragmentation. Also the radial
drift of particles trapped in the vortices is significantly reduced \citep{delaFuenteMarcosBarge2001}.
Theoretical attention has furthermore been given to the trapping of dust grains in high
pressure regions. Since
dust grains do not feel pressure forces, any pressure-supported gas structure
must cause dust grains to move in the direction of the pressure gradient
\citep{KlahrLin2001,HaghighipourBoss2003,KlahrLin2005}. Recently,
\cite{Rice+etal2004} demonstrated that this can lead to large concentrations (a
density increase of up to a factor 50) of meter-sized boulders in the high
density spiral arms of self-gravitating discs. The same mechanism can drain
millimeter-sized dust grains from the underdense regions around a protoplanet
that is not massive enough to open a gap in the gaseous component of the disc
\citep{PaardekooperMellema2004}.

\begin{deluxetable*}{crrrrrrr}

  \tablecaption{Simulation parameters}

  \tablewidth{0pt}

  \tablehead{
    \colhead{Run} & \colhead{$N$} & \colhead{$L_x\times L_y\times L_z$} &
    \colhead{$n_x\times n_y \times n_z$} & \colhead{$n_0$} &
    \colhead{$\varOmega_0 \tau_{\rm f}$} & \colhead{$\beta$} & \colhead{$\Delta
    t$} }

  \startdata
    A & $2\times10^6$ & $1.32\times 1.32\times1.32$ & $64\times 64\times64$ & $7.6$ & $ 1.0$ & $-0.04$ & $100$\\
    B & $2\times10^6$ & $1.32\times 1.32\times1.32$ & $64\times 64\times64$ & $7.6$ & $ 0.1$ & $-0.04$ & $100$\\
    C & $2\times10^6$ & $1.32\times 1.32\times1.32$ & $64\times 64\times64$ & $7.6$ & $10.0$ & $-0.04$ & $100$\\
    D & $2\times10^6$ & $1.32\times 1.32\times1.32$ & $64\times 64\times64$ & $7.6$ & $ 1.0$ & $ 0.00$ & $100$\\
    E & $2\times10^6$ & $1.32\times 5.28\times1.32$ & $64\times256\times64$ & $1.9$ & $ 1.0$ & $-0.04$ & $ 24$\\ 
    F & $2\times10^6$ & $1.32\times10.56\times1.32$ & $64\times512\times64$ & $1.0$ & $ 1.0$ & $-0.04$ & $ 16$\\
  \enddata

  \tablecomments{First column: name of run; second column: number of particles;
    third column: size of the box measured in scale heights; fourth column:
    grid dimension; fifth column: number of particles per grid cell; sixth
    column: friction time; seventh column: global pressure gradient parameter;
    eighth column: number of orbits that the simulation has run.}
  \label{t:simpar}

\end{deluxetable*}
Giant long-lived vortices may form in protoplanetary disc due to a baroclinic
instability \citep{KlahrBodenheimer2003}, but the conditions for the baroclinic
instability in protoplanetary discs are still not clear \citep{Klahr2004}. Magnetorotational
turbulence (MRI) on the other hand is expected to occur in all discs where the
ionization fraction is sufficiently high
\citep{Gammie1996,Fromang+etal2002,Semenov+etal2004}. A search for
dust concentrations in magnetorotational turbulence was done by
\cite{HodgsonBrandenburg1998} who found no apparent concentrations. On the
other hand, recently \citeauthor{JohansenKlahr2005} (2005, hereafter referred
to as JK05)
found evidence for centimeter-sized dust grains being trapped in short-lived
turbulent eddies present in magnetorotational turbulence. That work was,
however, limited by the fluid description of dust grains, i.e.\ the friction
time must be much shorter than the orbital period, and could not handle grains
larger than a few centimeters. 

In this paper, we expand the work done in JK05 by putting meter-sized dust
particles, represented by real particles rather than by a fluid, into
magnetorotational turbulence. We show that magnetorotational turbulence \citep{BalbusHawley1991}
is not actually an obstacle to the self gravity-aided formation of
planetesimals, but rather can be a vital agent to produce locally gravitational
unstable regions in the solid component of the disc when the average density in
solids would not allow for fragmentation. This process is very similar to the
gravoturbulent fragmentation of molecular clouds into protostellar cores
\citep{Klessen+etal2000,PadoanNordlund2004}.

\section{DYNAMICAL EQUATIONS}

For the purpose of treating meter-sized dust boulders we have adapted the
Pencil Code\footnote{The code is available at\\
\url{http://www.nordita.dk/software/pencil-code/}.} \citep[see also][]
{Brandenburg2003} to include the treatment of solid bodies as particles
with a freely evolving $(x,y,z)$-coordinate on top of the grid. This is
necessary because the mean free path of the boulders, with respect to
collisions with the gas molecules, is comparable to the scale height of the
disc. Thus the dust component can no longer be treated as a fluid, but must be
treated as particles each with a freely evolving spatial coordinate $\vc{x}_i$
and velocity vector $\vc{v}_i$. In other words, it is no longer possible to
define a unique velocity field at a given point in
space for the particles, because they keep a memory of their previous motion. 
Friction only erases this memory for small grains.
\\ \\

\subsection{Drag Force}

The particles are coupled to the gas motion by a drag force that is
proportional to the velocity difference between the particles and the gas,
\begin{equation}\label{eq:fdrag}
  \vc{f}_{\rm drag} = -\frac{1}{\tau_{\rm f}} (\vc{v}_i-\vc{u}) \, .
\end{equation}
Here $\vc{u}$ is the gas velocity at the location of particle $i$ and
$\tau_{\rm f}$ is the friction time. The friction
time depends on the solid radius $a_\bullet$ and the solid density
$\rho_\bullet$ as
\begin{equation}\label{eq:tauf}
  \tau_{\rm f} = \frac{a_\bullet^2 \rho_\bullet}{{\rm min}(a_\bullet c_{\rm
  s},\frac{9}{2} \nu) \rho} \, ,
\end{equation}
where $\nu$ is the molecular viscosity of the gas, $c_{\rm s}$ is the sound
speed and $\rho$ is the gas density. This expression is valid when the
particle speed is much lower than the sound speed \citep{Weidenschilling1977}.
Using the kinetic theory expression for viscosity $\nu=c_{\rm s} \lambda/2$,
where $\lambda$ is the mean free path of the gas molecules, 
the friction time can be divided into two regimes: the Epstein regime is valid when
$a_\bullet < 9/4 \lambda$. Here the mean free path of the gas molecules
is longer than the size of the dust grain, so the gas can not form any flow
structure around the object. The friction time is proportional to the
solid radius in this regime. In the Stokes regime, where $a_\bullet
> 9/4 \lambda$, a flow field forms around the object. Now the friction time is
proportional to solid radius squared, so the object decouples faster from the
gas with increasing size. For an isothermal and unstratified disc, one can
treat the friction time $\tau_{\rm f}$ as a constant. The distinction between
the Epstein and the Stokes regime is then only important for translating the
friction time into a solid radius (see end of this section).
\begin{deluxetable*}{crrrrrrrrr}

  \tablecaption{Results}

  \tablewidth{0pt}

  \tablehead{
    \colhead{Run} &
    \colhead{$\varOmega_0 \tau_{\rm f}$} & \colhead{$\beta$} &
    \colhead{${\rm max}(n)$} &
    \colhead{$\overline{v_x}$} &
    \colhead{$v_x^{\rm (lam)}$} & \colhead{$\overline{\sigma}$} &
    \colhead{$\overline{\sigma_x}$} & \colhead{$\overline{\sigma_y}$} &
    \colhead{$\overline{\sigma_z}$} }

  \startdata
    A & $ 1.0$ & $-0.04$ & $81.3$ & $-0.0123$ & $-0.020$ & $0.0222$ & $0.0162$ & $0.0105$ & $0.0077$\\
    B & $ 0.1$ & $-0.04$ & $32.6$ & $-0.0034$ & $-0.004$ & $0.0139$ & $0.0064$ & $0.0101$ & $0.0052$\\
    C & $10.0$ & $-0.04$ & $77.5$ & $-0.0042$ & $-0.004$ & $0.0170$ & $0.0115$ & $0.0094$ & $0.0062$\\
    D & $ 1.0$ & $ 0.00$ & $56.5$ & $-0.0003$ & $ 0.000$ & $0.0225$ & $0.0165$ & $0.0106$ & $0.0078$\\
    F & $ 1.0$ & $-0.04$ & $50.3$ & $-0.0132$ & $-0.020$ & $0.0204$ & $0.0149$ & $0.0093$ & $0.0066$\\
    E & $ 1.0$ & $-0.04$ & $50.3$ & $-0.0132$ & $-0.020$ & $0.0194$ & $0.0140$ & $0.0086$ & $0.0061$\\
  \enddata

  \tablecomments{First column: name of run; second column: friction time;
    third column: global pressure gradient parameter;
    fourth column: maximum particle
    density in units of the average density; fifth column: radial velocity
    averaged over space and time; sixth column: predicted radial drift in a
    non-turbulent disc; seventh to tenth columns: velocity dispersion averaged
    over space and time. Averages are taken from 5 orbits and beyond.  Grid
    cells with 0 or 1 particles have been excluded for the calculations of
    velocity dispersions.}
  \label{t:simres}

\end{deluxetable*}

To determine the gas velocity in \Eq{eq:fdrag} at the positions of the
particles, we use a three-dimensional first-order interpolation scheme, using
the eight grid corner points surrounding a given particle. For multiprocessor
runs the particles can move freely between the spatial intervals assigned to
each processor using MPI (Message Passing Interface) communication.

\subsection{Disc Model}

We consider a protoplanetary disc in the shearing sheet approximation, but
for a disc with
a radial pressure gradient $\dpa \ln P/\dpa \ln r = \alpha$ (or $P
\propto r^\alpha$). In the shearing sheet approximation this
gradient produces a constant additional force that points radially outwards
(because the pressure falls outwards). Making the variable transformation
$\ln\rho\rightarrow\ln\rho+(1/r_0)\alpha x$, the standard isothermal shearing
sheet equation of motion \citep[e.g.][]{GoldreichTremaine1978} gets an extra term,
\begin{equation}\label{eq:ss_simple}
  \frac{\dpa \vc{u}}{\dpa t} + (\vc{u}\cdot\nab)\vc{u}
      = -2\vc{\varOmega_0}\times\vc{u} + 3\varOmega_0^2 \vc{x}
      -c_{\rm s}^2 \nab \ln \rho - c_{\rm s}^2 \frac{1}{r_0} \alpha \hat{\vc{x}}
  \, .
\end{equation}
The terms on the right-hand-side of \Eq{eq:ss_simple} are the Coriolis force,
the centrifugal force plus the radial gravity expanded to first order, and the
two terms representing local and global pressure gradient.
The coordinate vector $(x,y,z)$ is measured from the comoving radial position
$\vc{r}_0$ from the central source of gravity, with $x$ pointing radially
outwards and $y$ along the Keplerian flow. At $r=r_0$ the Keplerian
frequency is $\varOmega_0$. The shearing sheet approximation is valid when
all distances are much shorter than $r_0$. The balance between pressure
gradient, centrifugal force and gravity is given for a sub-Keplerian rotation
of the disc, 
\begin{equation}
  u_y^{(0)} = -\frac{3}{2} \varOmega_0 x
      + \frac{c_{\rm s}^2}{2 \varOmega_0} \frac{1}{r_0} \alpha \, ,
\end{equation}
where the first term on the right-hand-side is the purely Keplerian rotation
profile, while the second (constant) term is the adjustment due to the global
pressure gradient. We now measure all velocities relative to the sub-Keplerian
flow using the variable transformation $\vc{u}\rightarrow\vc{u}+\vc{u}_0$. This
changes \Eq{eq:ss_simple} into
\begin{equation}
  \frac{\dpa \vc{u}}{\dpa t} + (\vc{u}\cdot\nab)\vc{u}
      + u_y^{(0)}\frac{\dpa \vc{u}}{\dpa y}
      = \vc{f}(\vc{u}) -c_{\rm s}^2 \nab \ln \rho \, .
\end{equation}
Here the last term on the left-hand-side represents the advection due to the 
rotation of the disc relative to the center of the box (which moves on a purely
Keplerian orbit).
The function $\vc{f}$ is defined as
\begin{equation}
  \vc{f}(\vc{u}) = \pmatrix{2 \varOmega_0 u_y \cr -
      \frac{1}{2} \varOmega_0 u_x \cr 0} \, .
\end{equation}
When making the same variable transformation in the equation of motion of the
dust particles, there is however no global pressure gradient term to balance
the extra Coriolis force imposed by the sub-Keplerian part of the motion, so
the result is
\begin{equation}\label{eq:dvdt_gpr}
  \frac{\dpa \vc{v}_i}{\dpa t} = \vc{f}(\vc{v}_i)
      - \frac{1}{\tau_{\rm f}}(\vc{v}_i - \vc{u})
      + c_{\rm s}^2 \frac{1}{r_0} \alpha \hat{\vc{x}} \, .
\end{equation}
The modified Coriolis force $\vc{f}$ appears again because of the presence
of $x_i(t)$ in $\vc{u}_0$. The last term on the right-hand-side reflects the
head wind that the dust feels when it moves through the slightly sub-Keplerian
gas. The reason that the term appears in the radial component of the equation
of motion is that all velocities are measured relative to the rotational
velocity of the gas. A dust particle moving at zero velocity with respect to
the gas thus experiences an acceleration in the radial direction.
\begin{figure*}
  \includegraphics{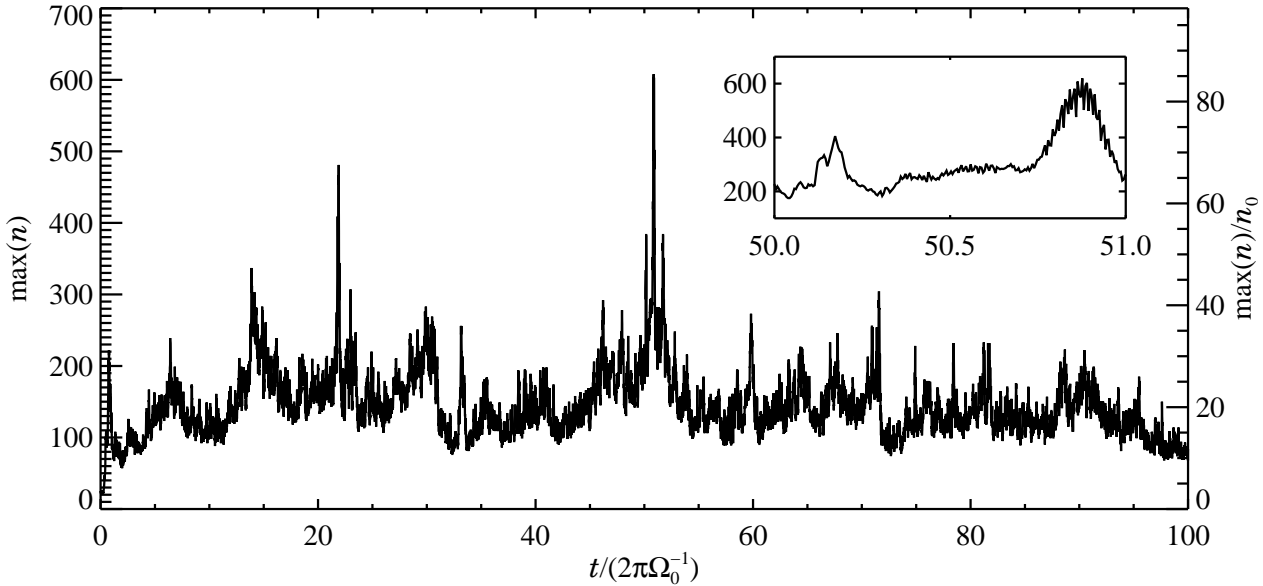}
  \caption{The number of particles in the densest grid cell as a function of
    time for run A (meter-sized boulders). The maximum density is generally
    around 20 times the average, but peaks at above 80 times the average
    particle density. The insert shows a magnification of the time between 50
    and 51 orbits.}
  \label{f:npmax_t}
\end{figure*}

The explicit presence of $r_0$ in \Eq{eq:dvdt_gpr} is non-standard in the
shearing sheet. It may seem that the term vanishes for $r_0\rightarrow\infty$.
But this is actually not the case, since the natural timescale of the disc,
$\varOmega_0^{-1}$, also depends on $r_0$, so that at large radii there is an
immense amount of time at hand to let the tiny global pressure gradient force
work. One can quantify this statement by dividing and multiplying by the
scale-height $H$ in the last term of \Eq{eq:dvdt_gpr} to obtain the result
\begin{equation}
  \frac{\dpa v_x^{(i)}}{\dpa t} =
      \ldots + c_{\rm s} \varOmega_0 \frac{H}{r_0} \alpha \, .
  \label{eq:headwind}
\end{equation}
Here $H/r_0\equiv\xi$ is the ratio of the scale height to the orbital radius,
a quantity that is below unity for thin discs. Depending on the temperature
profile of a disc, the typical value of $\xi$ is between 0.001 and 0.1. We
define the pressure gradient parameter $\beta$ as $\beta \equiv \alpha \xi$.

\begin{figure*}
  \includegraphics{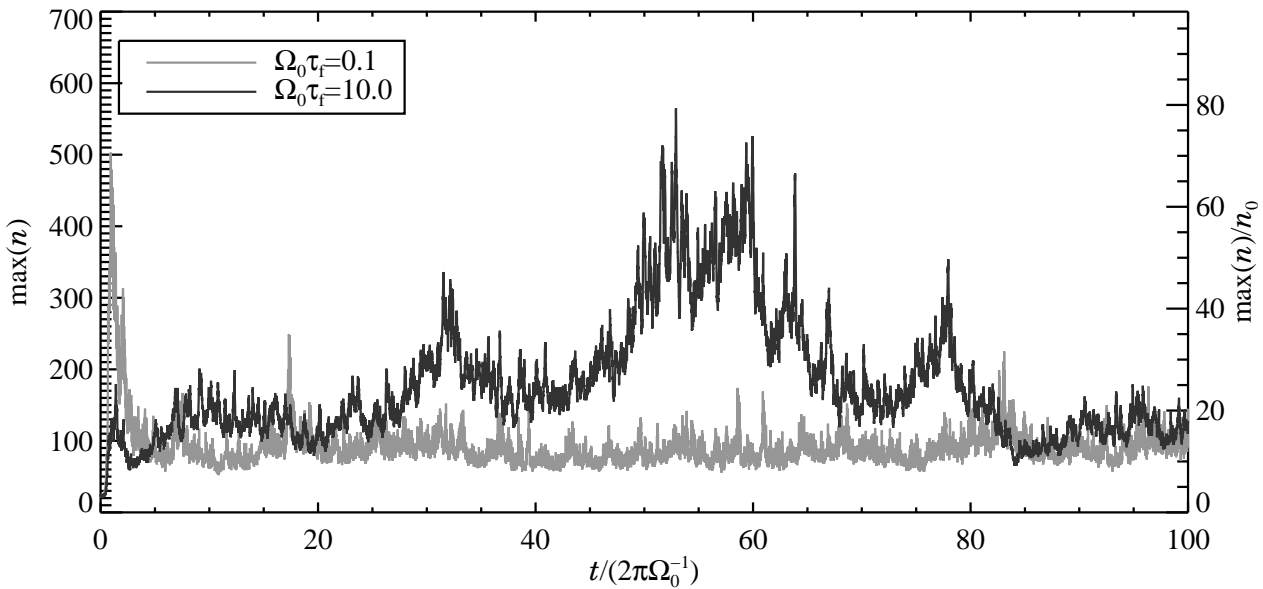}
  \caption{The number of particles in the densest grid cell as a function of
    time, here for runs B (decimeter-sized boulders) and C (decameter-sized
    boulders). The first shows only very moderate overdensities, whereas the
    latter is similar in magnitude to run A (meter-sized boulders), but with
    broader peaks.}
  \label{f:npmax_t2}
\end{figure*}
For the simulations, we adopt the following dynamical equations for gas
velocity $\vc{u}$, magnetic vector potential $\vc{A}$, gas density $\rho$, 
particle velocities $\vc{v}_i$ and particle coordinates $\vc{x}_i$:
\begin{eqnarray}
  \frac{\dpa \vc{u}}{\dpa t} + (\vc{u}\cdot\nab)\vc{u}
      + u_y^{(0)} \frac{\dpa\vc{u}}{\dpa y} &=& \vc{f}(\vc{u})
      - c_{\rm s}^2 \nab \ln \rho \nonumber \\
      && \hspace{0.2cm} + \frac{1}{\rho} \vc{J} \times \vc{B}
      + \vc{f}_\nu (\vc{u},\rho)
      \label{eq:dyneq_gas}\\
  \frac{\dpa \vc{A}}{\dpa t} + u_y^{(0)} \frac{\dpa\vc{A}}{\dpa y} &=&
      \vc{u} \times \vc{B} \nonumber \\
      && \hspace{0.2cm} + \frac{3}{2} \varOmega_0 A_y \hat{\vc{x}} + 
      \vc{f}_\eta(\vc{A}) \\
  \frac{\dpa \rho}{\dpa t} + \vc{u} \cdot \nab \rho  +
       u_y^{(0)} \frac{\dpa \rho}{\dpa y} &=&
       -\rho \nab \cdot \vc{u} + f_{\rm D}(\rho)
      \label{eq:dyneq_dens}\\
  \frac{\dpa \vc{v}_i}{\dpa t}
      &=& \vc{f}(\vc{v}_i) + c_{\rm s} \varOmega_0 \beta \hat{\vc{x}}
      \nonumber \\
      && \hspace{0.2cm} - \frac{1}{\tau_{\rm f}} (\vc{v}_i-\vc{u})
      \label{eq:dyneq_par_v}\\
  \frac{\dpa \vc{x}_i}{\dpa t}
      &=& \vc{v}_i + u_y^{(0)} \hat{\vc{y}}
      \label{eq:dyneq_par_x}
\end{eqnarray}
The functions $\vc{f}_\nu$, $\vc{f}_\eta$ and $f_{\rm D}$ are hyperdiffusivity
terms present to stabilize the finite difference numerical scheme of the Pencil
Code. This is explained in more detail in JK05. We shall ignore the effect of
the global pressure gradient on the dynamics of the gas, since for $\xi \ll 1$
the increase in density due to the global gradient is much smaller than the
average density in the box. Thus we set simply $u_y^{(0)}=-3/2 \varOmega_0 x$.
We also ignore the contribution from the global density on the Lorentz force term in
\Eq{eq:dyneq_gas} and the advection of global density in \Eq{eq:dyneq_dens}.
Furthermore we do not include vertical gravity in the simulations.
This means that we solve exactly the same equations for the gas as in JK05,
i.e.\ without radial pressure stratification.
The radial drift of solids then originates exclusively from the
dynamical equations of the particles.

We solve the dynamical \Eqss{eq:dyneq_gas}{eq:dyneq_par_x} for various values of the
friction time and of the box size. The typical resolution is $64^3$ for a box
size of $1.32 H$ on all sides.  A similar setup was
used in JK05 to calculate the turbulent diffusion coefficient of dust
grains in magnetorotational turbulence. In the present work we expand the model by letting 2,000,000 particles
represent the dust grains. Thus the dust component is typically represented by
approximately 8 particles per grid cell. We set the strength of the radial pressure
gradient by the parameter $\beta=-0.04$. This would represent e.g.\ a disc with
a global pressure gradient given by $\alpha=-1$ and a scale-height-to-radius
ratio of $\xi=0.04$, which is typical for a solar nebula model
\citep{WeidenschillingCuzzi1993}.
We consider friction times of $\varOmega_0\tau_{\rm
f}=0.1,1,10$. The
translation from friction time into grain size depends on whether the friction
force is in the Epstein or in the Stokes regime, but 
the two drag laws
yield quite similar grain sizes in the transition regime. Thus, at the radial location of Jupiter in a
typical protoplanetary disc, the friction time corresponds to grains of
approximately 0.1, 1 and 10 meters in size.

The simulation parameters are given in \Tab{t:simpar}. We let the boulders
have random initial positions from the beginning and let them start with zero
velocity.
\\ \\

\section{PARTICLE CONCENTRATIONS}

\begin{figure*}
  \includegraphics{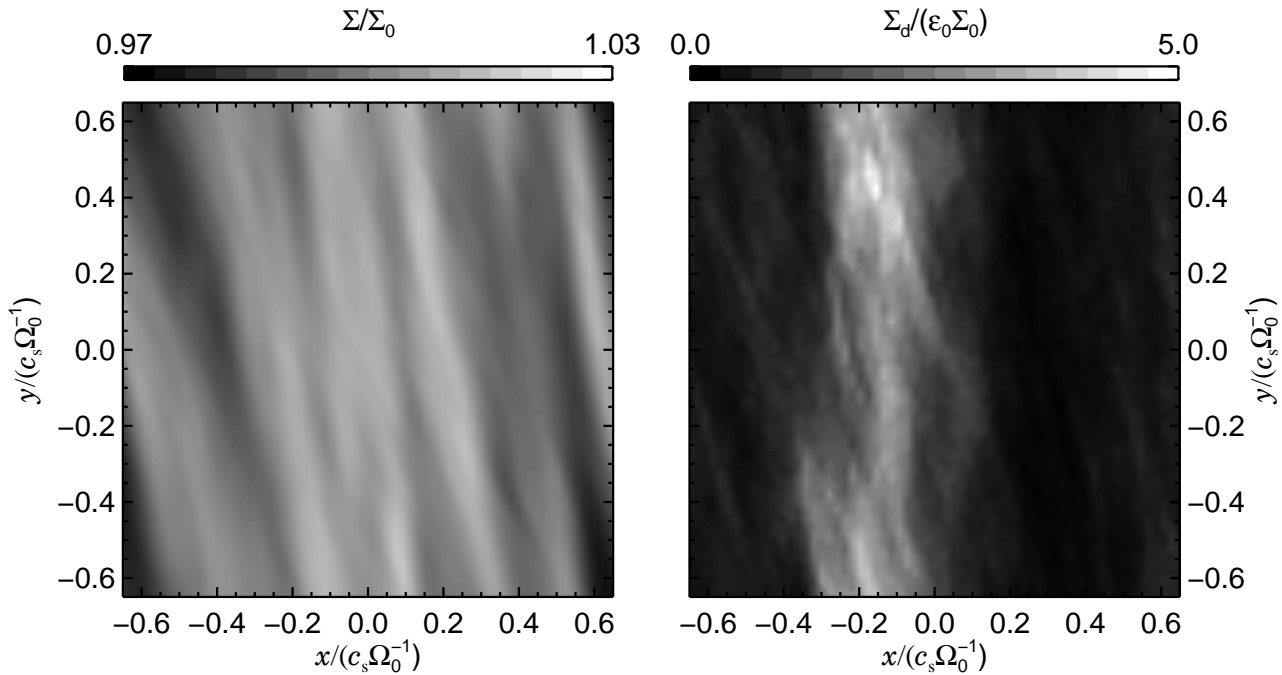}
  \caption{Gas column density $\varSigma$ (left panel) and dust column density
    $\varSigma_{\rm d}$ (right panel). The gas column density only varies by a
    few percent over the box, but still a slightly overdense region is seen near
    the center of the box. The dust column density in the same region is up to
    5 times the average dust column density.}
  \label{f:Sigma_Sigmad}
\end{figure*}
In \Fig{f:npmax_t} we plot the number of particles in the densest grid cell as
a function of time for run A (meter-sized boulders, see Table \ref{t:simpar}).
The average number of particles per grid cell is
$7.6$. Evidently there is more than 100 particles in the densest grid cell at
most of the times, and at some times the number is even above 600.
This is more than 80 times the average dust number density.
In \Fig{f:npmax_t2} we plot the
maximum particle density for runs with $\varOmega_0 \tau_{\rm f}=0.1$ (run B,
gray curve) and $\varOmega_0 \tau_{\rm f}=10$ (run C, black curve). The
decimeter-sized boulders are obviously not as strongly concentrated as the
meter-sized boulders, whereas the decameter-sized boulders have concentrations
that are similar in magnitude to run A.
The measured values of the maximum particle
density for all the runs can be found in \Tab{t:simres}.

To examine whether some structures in the gas density are the source of the high particle densities,
we plot in \Fig{f:Sigma_Sigmad} the column densities of gas $\varSigma$ and of
dust particles $\varSigma_{\rm d}$ at a time of 50.9 orbits for run A. The gas
column density varies only by a few percent over the box, since the turbulence
is highly subsonic, but a region of moderate overdensity is seen around the
middle of the box. The dust column density is very high in about the same
region as the gas overdensity, around a factor of five higher than the average
dust column density in the box, so dust particles have moved from the regions
that are now underdense into the overdensity structure near the center of the
box.

\begin{figure*}
  \includegraphics{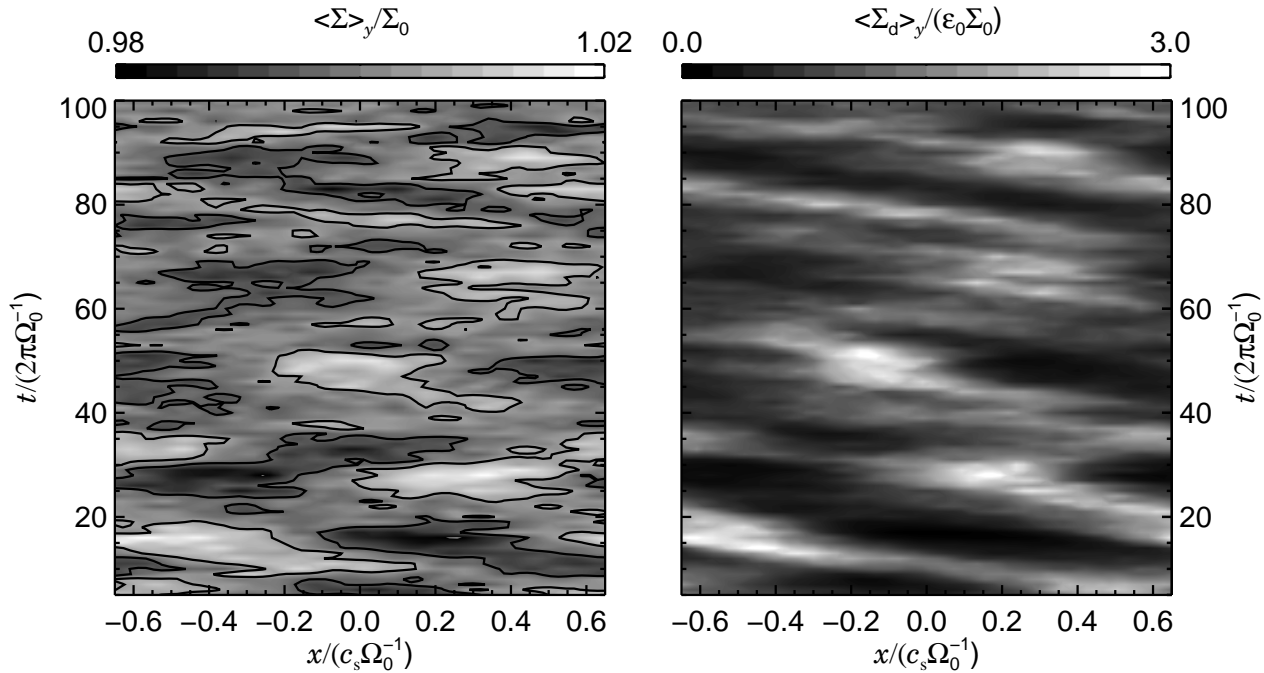}
  \caption{Azimuthally averaged gas and dust column densities as a function of
    radial position relative to the center of the box $x$ and time $t$. Black
    contour lines are shown at gas density fluctuations of $0.5\%$ from the
    average value. Large scale density fluctuations are seen to have lifetimes
    on the order of a few orbits before moving to other radial positions. The
    dust column density peaks strongly at the locations of the maximal gas
    column density.}
  \label{f:rhox_vs_t}
\end{figure*}

We explore the radial density structure of the gas and the dust in the box in more
detail in \Fig{f:rhox_vs_t}. Here the azimuthally averaged gas and dust column
densities are shown as a function of radial position $x$ and time $t$ measured in orbits.
Apparently large scale gas density fluctuations live for a few orbits at a
constant radial position before decaying and reappearing at another radial
position. The fluctuation strength is less than 1\% of the average density. The
dust density shows strong peaks at the locations of the gas density maxima. The
explanation for this correlation is as follows.
Locations of maximal gas density are also local pressure maxima. Such pressure
maxima can trap dust grains \citep{KlahrLin2001,HaghighipourBoss2003} as
they are locations of Keplerian gas motion. The inner edge of a pressure
maximum must move faster than the Keplerian speed because the pressure gradient
mimics an additional radial gravity. At the outer edge of a radial pressure
enhancement the outwards-directed pressure gradient mimics a decreased gravity,
and the gas must move slower than the Keplerian speed. Dust grains do not
feel the pressure gradient and are thus forced to move into the pressure bump.
In our simulations the radial gas overdensities have a typical lifetime at a
given radial position on the order of a few orbits. When the gas overdensity
eventually disappears, the particle overdensity is only slowly getting
dissolved, and the particles drift and concentrate towards the location of the
next gas overdensity. The gas density structure in the azimuthal and vertical
directions does not show a similar density increase, and as expected there is
also no significant concentration of particles with respect to these two
directions. The density fluctuations thus have the form of two-dimensional
sheets.

\begin{figure}
  \includegraphics{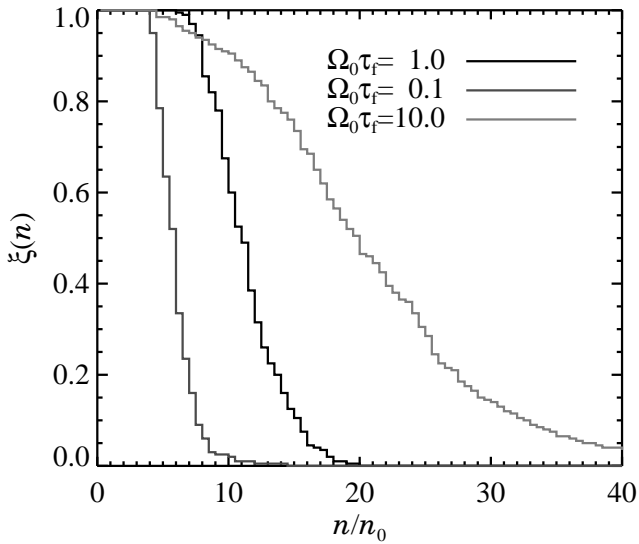}
  \caption{Distribution of maximum particle densities. The curves show the
    fraction $\xi(n)$ of particles that have been part of a given particle
    density during the 100 orbits. For $\varOmega_0 \tau_{\rm f}=0.1$, only
    concentrations up to 10 are common, whereas for $\varOmega_0 \tau_{\rm
    f}=1$, 70\% of the particles have experienced at least a 10 times increase
    in density and 2\% even a 20 times increase. For massive boulders with
    $\varOmega_0 \tau_{\rm f}=10$, more than 10\% were part of a 30 times
    increase in density. }
  \label{f:npdistr}
\end{figure}

In \Fig{f:npdistr} we plot the maximum density experienced by a 200 particle
subset of the 2,000,000 particles during the 100 orbits. The distribution
function $\xi(n)$ is defined as the fraction of particles that have been the
center of a number density of at least $n$ over the size of a grid cell. The curves
clearly show how large the concentrations are. For decimeter-sized
boulders, 95\% of them have experienced a 5 times increase in dust
density, whereas only around 2\% have been part of a 10 times increase. For
meter-sized particles, 70\% have been part of a 10 times increase in dust
density, and 1\% even took part in a 20 times increase. Particles of
decameter-size had more than 10\% taking part in a 30 times increase of dust
density. This is very similar to the concentrations that \cite{Rice+etal2004}
find in the spiral arms of self-gravitating discs.

\begin{figure*}
  \includegraphics{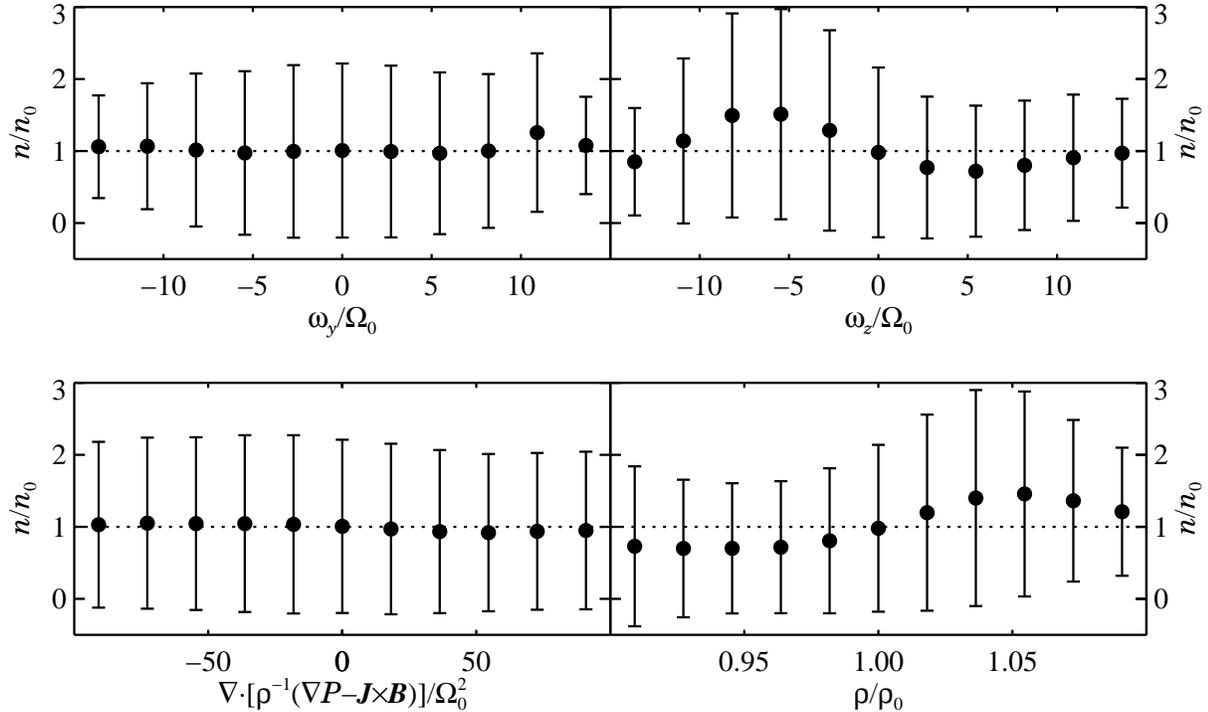}
  \caption{Correlations between particle number density and various gas
    parameters. The first row considers two components of the vorticity. There
    is some correlation between $n$ and $\omega_z$, indicating that particles
    are trapped in regions of anticyclonic flow. In the second row we consider
    the correlation between particle density and pressure gradient flux
    (explained in the text) and gas density, respectively. The first
    correlation is very weak, whereas there is evidently a correlation between
    gas density and particle density, although the fluctuation bars are
    significant.}
  \label{f:np_misc}
\end{figure*}

In \Fig{f:np_misc} correlations between gas flow and particle density
are shown for run D (without global pressure gradient). Here we have taken data at every full orbit, starting at 5 orbits
when the turbulence has saturated, and calculated the average particle density
in bins of various gas parameters. We also plot the spread in the particle
density in each bin. The top two panels show the correlation with two
components of the vorticity $\vc{\omega}=\nab\times\vc{u}$. There is some
correlation between
vertical vorticity component and the particle density, but the spread in each
vorticity bin is larger than the average value. The correlation indicates that some
trapping of particles is happening in anticyclonic regions, and that regions of
cyclonic flow are expelling particles \citep{BargeSommeria1995}. Meter-sized
particles should be optimally concentrated by vorticity, so the weak
correlation between $n$ and $\omega_z$ is surprising, considering that for
centimeter-sized particles, JK05 find an almost linear relation between $n$
and $\omega_z$ with very small spread. The explanation may be that the friction
time is so high that particle concentrations stay together even after the gas
feature which created them has decayed or moved to another location. The
limited life-time of the concentrating features weakens the measured
correlation with the gas flow.

The lower two
panels of \Fig{f:np_misc} show the correlation with divergence of pressure gradient flux and with
gas density.
In a steady flow, particles accelerate towards an equilibrium velocity where the drag force is
in balance with the other forces working on the particles. The equilibrium
velocity is
\begin{equation}\label{eq:sfta}
  \vc{v} = \tau_{\rm f} \rho^{-1} \left( \nab P - \vc{J} \times \vc{B} \right)
      \equiv \tau_{\rm f} \vc{F} \, .
\end{equation}
This is the mechanism for pressure gradient-trapping. Places with a
negative value of $\nab\cdot\vc{F}$
should produce
a high particle density (see JK05). The correlation between
$\nab\cdot\vc{F}$ and $n$ is existent, but is very weak. The last panel, however, shows
that there is a clear correlation between gas density and particle density, as
is also evident from \Fig{f:rhox_vs_t}.
All in all, the correlations, even though some of the are quite weak, give the necessary information about the source of
the dust concentrations. The concentrations are primarily due to pressure
gradient-trapping in the gas flow. There is also evidence of some
vorticity-trapping happening on top of that.

Increases in density of up to two orders of magnitude will make a difference in
the coagulation process, because at places of larger concentration more
collisions (both destructive and constructive) are possible. Also there is a chance of increasing the density to
such
high values that a gravitational instability can occur in the densest places. We
will consider this last point in more detail in \Sec{ch:gi}. In the following
section we show that the turbulence not only causes concentrations, but also
changes the radial drift velocity of the boulders.

\section{DRIFT SPEED}\label{ch:driftspeed}

The global pressure gradient on the gas forces solids to fall radially inwards.
If the gas motion in the disc was completely non-turbulent, then the
equilibrium radial drift velocity arising from the head wind term present in
\Eq{eq:dyneq_par_v} would be
\begin{equation}\label{eq:vxdrift}
    v_x = \frac{\beta}{\varOmega_0 \tau_{\rm f}+(\varOmega_0 \tau_{\rm f})^{-1}}
        c_{\rm s} \, .
\end{equation}
We derived this expression by solving for $\dpa \vc{v}/\dpa t=0$ in
\Eq{eq:dyneq_par_v}.
The highest drift speed occurs for particles with
$\varOmega_0 \tau_{\rm f}=1$ with a laminar drift velocity of $v_x/c_{\rm s} =
\beta/2$. We have checked by putting particles of different friction times into
a non-turbulent disc that the measured drift velocities are in complete
agreement with \Eq{eq:vxdrift}.

\begin{figure*}
  \includegraphics{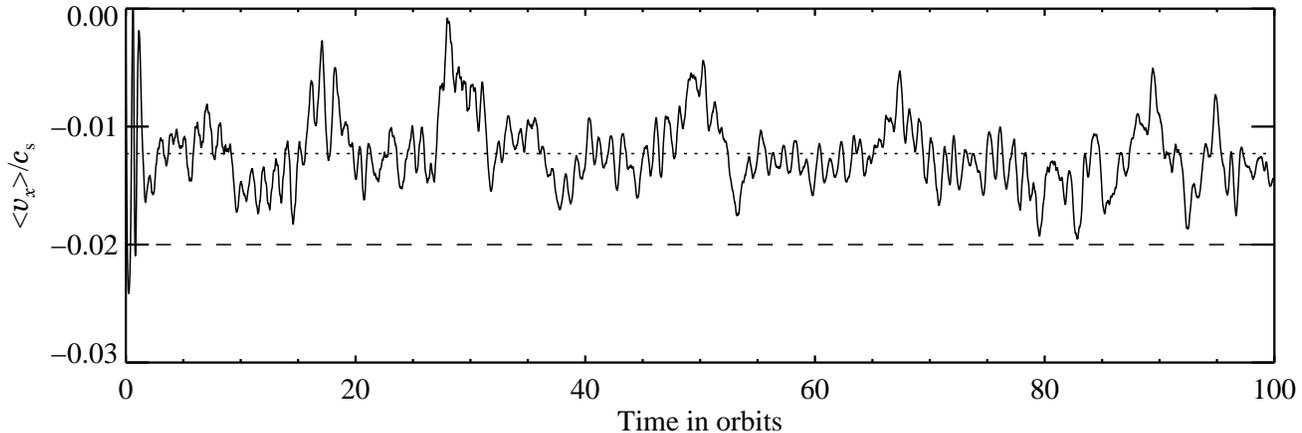}
  \caption{Average radial particle velocity as a function of time for
    meter-sized boulders. The non-turbulent drift velocity is $v_x^{\rm
    (lam)}=-0.02\,c_{\rm s}$ (indicated with a dashed line), while the average
    drift velocity in the turbulent case is only around
    $\overline{v_x}=-0.012\,c_{\rm s}$, a reduction by around $40\%$ in
    speed.}
  \label{f:vpxm_vs_time}
\end{figure*}

\begin{figure}
  \includegraphics[width=\linewidth]{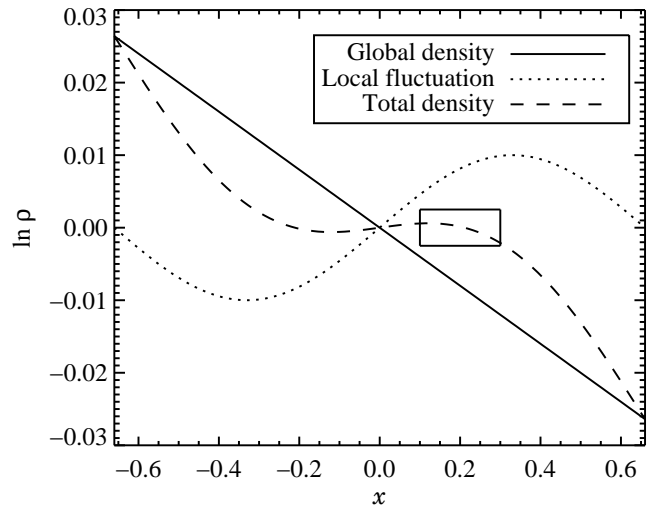}
  \caption{Sketch of how turbulent density fluctuations can cause the average
    drift velocity to change. The full line shows the global density as a
    function of radial distance from the center of the box. On top of this we
    sketch a large-scale sinusoidal density fluctuation (dotted line) and the
    total density (dashed line). Dust particles are concentrated in the positive part of the
    fluctuation. At the same time the concentration drifts towards the location of the box where the
    total drift speed is zero. If a significant fraction of the particles
    end up in such regions, then the average drift speed can decrease.}
  \label{f:trapping}
\end{figure}

The effect of a real turbulent disc on the average drift velocity is seen in
\Fig{f:vpxm_vs_time}. Here the average radial velocity of all the
particles is shown as a function of time for run A.
For reference we overplot the laminar drift velocity ($v_x=-0.02\,c_{\rm s}$) from
\Eq{eq:vxdrift} and the time-averaged drift velocity
($\overline{v_x}=-0.012\,c_{\rm s}$). The
mean drift velocity is noticeably affected by the turbulence and its absolute
value is reduced by 40\% compared to the laminar value. 
The influence that turbulence can have on the mean drift
velocity of the particles can be quantified with some simple analytical
considerations. Considering the particles for a moment as a fluid with a number
density scalar field $n$ and a velocity vector field $\vc{w}$, the average
radial velocity can be calculated with the expression
\begin{equation}\label{eq:wx_ave}
  \langle w_x \rangle = \frac{\int_{x_0}^{x_1} n w_x \de x}{\langle n \rangle
  L_x}
  \, .
\end{equation}
Here we have weighted the drift velocity with the number density so that we are
effectively measuring the average momentum. We consider now for simplicity
particles that have been accelerated by the gas to their terminal velocity
(eq. [\ref{eq:vxdrift}] including the fluctuation pressure gradient),
\begin{equation}\label{eq:wx_sfta}
  w_x = \epsilon c_{\rm s} \left( \beta + \frac{\dpa \ln \rho}{\dpa x} \right) \, ,
\end{equation}
where $\epsilon$ is defined as $\epsilon=1/[\varOmega_0 \tau_{\rm f}+(\varOmega_0
\tau_{\rm f})^{-1}]$.
Inserting now \Eq{eq:wx_sfta} into \Eq{eq:wx_ave}, the resulting drift velocity
is found to consist of two terms,
\begin{equation}\label{eq:wx_ave2}
  \langle w_x \rangle = \epsilon \beta c_{\rm s} + 
      \frac{\epsilon c_{\rm s} \int_{x_0}^{x_1} n \frac{\dpa\ln\rho}
      {\dpa x} \de x}{\langle n \rangle L_x} \, .
\end{equation}
The first term on the right-hand-side of \Eq{eq:wx_ave2} represents the
contribution to the average drift velocity from the global pressure gradient
(eq.[\ref{eq:vxdrift}]). The other term is an extra contribution due to any
non-zero correlation between number density $n$ and radial pressure gradient
$\dpa \ln \rho/\dpa x$. This situation is sketched in \Fig{f:trapping}.  Here
we sketch the global density gradient $\beta$ (full line) and a sinusoidal
density fluctuation $\ln \rho (x)$ (dotted line). Particles concentrate
in regions where the gas density fluctuation is positive, because there the
divergence of the particle velocity is negative. Due to the total pressure
gradient, the newly produced particle clumps drift inwards until the point where
the outwards drift towards the fluctuation density maximum balances the inwards
drift from the global pressure gradient. This is exactly around the location of
the box in \Fig{f:trapping}. Here the correlation between $n$ and $\dpa \ln
\rho/\dpa x$ leads to a positive value of the integral in \Eq{eq:wx_ave2}. A
closer inspection of \Fig{f:rhox_vs_t} reveals that the dust overdensities are
situated slightly downstream of the gas density fluctuation peaks, which is in
good agreement with the the prediction in \Fig{f:trapping}. If a
significant fraction of the particles end up in such regions, the average drift
speed is reduced\footnote{A more graphic explanation of the speed
reduction is to consider a car race over a distance of 100 km.
Half of the distance is sand, where the cars can run 50 kilometers per hour,
and the other half asphalt, where the cars go 150 kilometers an hour. The
average speed of a single car reaching the finish line is less than 100
kilometers per hour, simply because that car spent more time on sandy terrain
than on asphalt.}.
For runs B and C, there is no significant reduction of the drift speed (see
Table \ref{t:simres}), but there the predicted drift speed is also ten times
lower than for meter-sized objects. Thus the measurement is not as reliable
because the random velocity fluctuations of the particles dominate over the
radial drift.

\begin{figure}
  \includegraphics{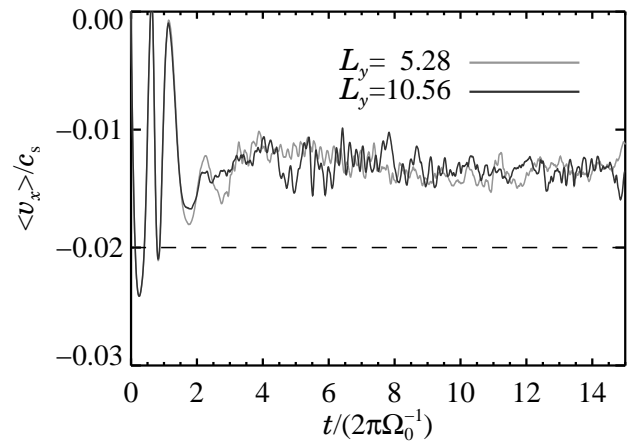}
  \caption{Drift velocity for simulations E and F with larger $y$-domains. The
    expected drift velocity in a laminar disc is indicated with a dashed line.
    The
    measured drift velocity is approximately the same as for the cube
    simulations, so the periodic $y$-boundary is not the reason for the reduced
    drift speed. Rather it is a side-effect of trapping the particles in radial
    density enhancements.}
  \label{f:vpxm_t2}
\end{figure}

Due to the periodic boundary conditions in the $y$-directions, density
structures quickly pass the $y$-boundaries, by shear advection, and thus
possibly have some interference with themselves. To see the effect of the
toroidal box size on the radial drift, we have run simulations with a box
size of $1.32\times5.28\times1.32$ (run E) and $1.32\times10.56\times1.32$
(run F), keeping the
resolution constant by adding the appropriate number of grid points in the
$y$-direction. The time evolution of the mean radial drift velocity is shown in \Fig{f:vpxm_t2}. It
is evidently very similar to \Fig{f:vpxm_vs_time}, so the toroidal size of the
box does not influence the radial drift reduction noticeably. As seen in
\Tab{t:simres}, the maximum particle density for runs E and F is quite high at
50 times the average density in the box, but not as high as in run A.
However, simulations E and F only ran for 24 and 16 orbits, respectively,
because of computational requirements due to the many grid points.

In simulations of the interaction between a planet and a magnetorotationally
turbulent disc, \cite{NelsonPapaloizou2004} find that the average migration
velocity of the planets is not changed by the presence of MRI turbulence
(whereas the spread in drift velocity causes some planets to even drift
outwards). On the other hand, recent simulations by \cite{Nelson2005} indicate
that the mean migration of planets can indeed change because of turbulence. The
fluctuations in migration speed are however much stronger than the average (so
that hundreds of orbits are needed for a reliable estimate of the average).
This is a very different kind of drift behavior than for the boulders in the
current work, where the fluctuations in the drift speed are actually much {\it
smaller} than the average. The presence of long-lived attracting regions in the
gas may be the reason why boulders react on turbulence in a completely different
way than planets do.

Diminishing the radial drift for meter-sized objects by roughly one half may not be
saving the boulders from their fate of decaying into the star. One will have to
investigate this process by additionally looking at the growth behavior of the
boulders which are sweeping up small grains on their way inwards. This sweeping
up is determined by the actual drift speed with respect to the local gas
motion. Even if the mean drift speed is
above the threshold for effective sticking, there will be phases of much lower
radial drift, where growth can occur. The overdense regions would also greatly
increase the rate of destructive encounters between larger bodies, and thus the
reservoir of small bodies would be stronger replenished there. This would not
only influence growth of larger bodies, but also possibly have observational
consequences.

The present simulations are done in the gentle situation of turbulence in a
local box. Global disc simulations have stronger turbulence and larger density
fluctuations. One can predict that it would thus also lead to a larger decrease
in radial drift speed. This would possibly give the meter-sized boulders
enough time to grow to a size safe for radial drift. However, this yet has to
be demonstrated in global simulations\footnote{We have recently become aware
of work done by \cite{FromangNelson2005} where the dynamics of boulders in
magnetorotational turbulence is considered in global simulations of accretion
discs. They found indeed that solids can be trapped inside persistent flow
features for even a hundred orbits, i.e.\ the entire simulation length.}.

\section{GRAVITATIONAL INSTABILITY}\label{ch:gi}

We already showed that turbulence can strongly influence the growth of
boulders by slowing them down and by concentrating them locally. These results
can be incorporated into standard evolution codes for the solid material
\citep[e.g.][]{Weidenschilling1997,DullemondDominik2005}, which try to grow
planetesimals from dust grains via coagulation. On the other hand the high
local concentration can also lead to a different way of planetesimal formation,
i.e.\ gelation. In the gelation case a cloud of boulders is so dense that
gravitational attraction becomes important. While we will not study self-gravity
by an $N$-body approach in this work (as one should), we want at least
demonstrate by simple estimations under what conditions the concentration of
boulders could clump into planetesimals.

The gravity constant $G$ enters in self-gravity calculations, and thus the
equations are no longer scale-free, but depend on the adopted disc model. We
characterize a disc model by a column density $\varSigma_0$, an average
dust-to-gas mass density ratio $\epsilon_0$ (for boulders of the considered
size range) and a scale-height-to-radius ratio of $\xi$. Of course,
$\epsilon_0$ will be smaller than the global dust-to-gas ratio $\sim0.02$,
because only a part of the mass will be present in boulders of the considered
size range. We choose for simplicity the value $\epsilon_0=0.01$, assuming that
50\% of the total dust mass is in bodies of the considered size, and we shall
later discuss in how far this value is reasonable for a protoplanetary disc.

The apparently large number of particles
in our numerical simulations is still orders of magnitude away from any real
number of boulders in the volume of the protoplanetary disc considered in our
simulations. Thus it is
necessary and validated to let one superparticle represent an entire swarm of
many particles of similar location and velocity in the disc. Superparticle
means in this context that one particle has the aerodynamic behavior of a
single boulder, but represents a mass of trillions of such bodies as it mimics
an entire swarm of protoplanetesimals. Similar assumptions are common in
simulations of giant planet core formation from colliding planetesimals
\citep{KokuboIda2002,Thommes+etal2003} as well as in cosmological
$N$-body simulations \citep{Sommer-Larsen+etal2003}. We let the simulation box
represent the protoplanetary disc in the mid-plane. Each superparticle then
contains the mass $m=\epsilon_1 \rho_1 V/N$, where $V$ is the volume of the
box, $N$ is the number of superparticles, and $\epsilon_1$ and $\rho_1$ are the
dust-to-gas ratio and the gas density in the mid-plane of the disc. We shall
use the isothermal disc expression $\rho_1=\varSigma_0/(\sqrt{2\pi}H)$ to
calculate the mass density in the mid-plane.

To calculate the dust-to-gas ratio in the mid-plane, $\epsilon_1$, one needs to
take into account the effect of vertical settling of solid material.
Solids move in the direction of higher gas pressure. In the
case of vertical stratification, that means that the boulders must sediment
towards the mid-plane.  An equilibrium is reached when the sedimentation is
balanced by the turbulent diffusion, with diffusion coefficient $D_{\rm t}$
\citep{SchraeplerHenning2004}, away from the mid-plane. This leads to a
Gaussian profile of the dust-to-gas ratio \citep{Dubrulle+etal1995}, 
\begin{figure*}
  \includegraphics{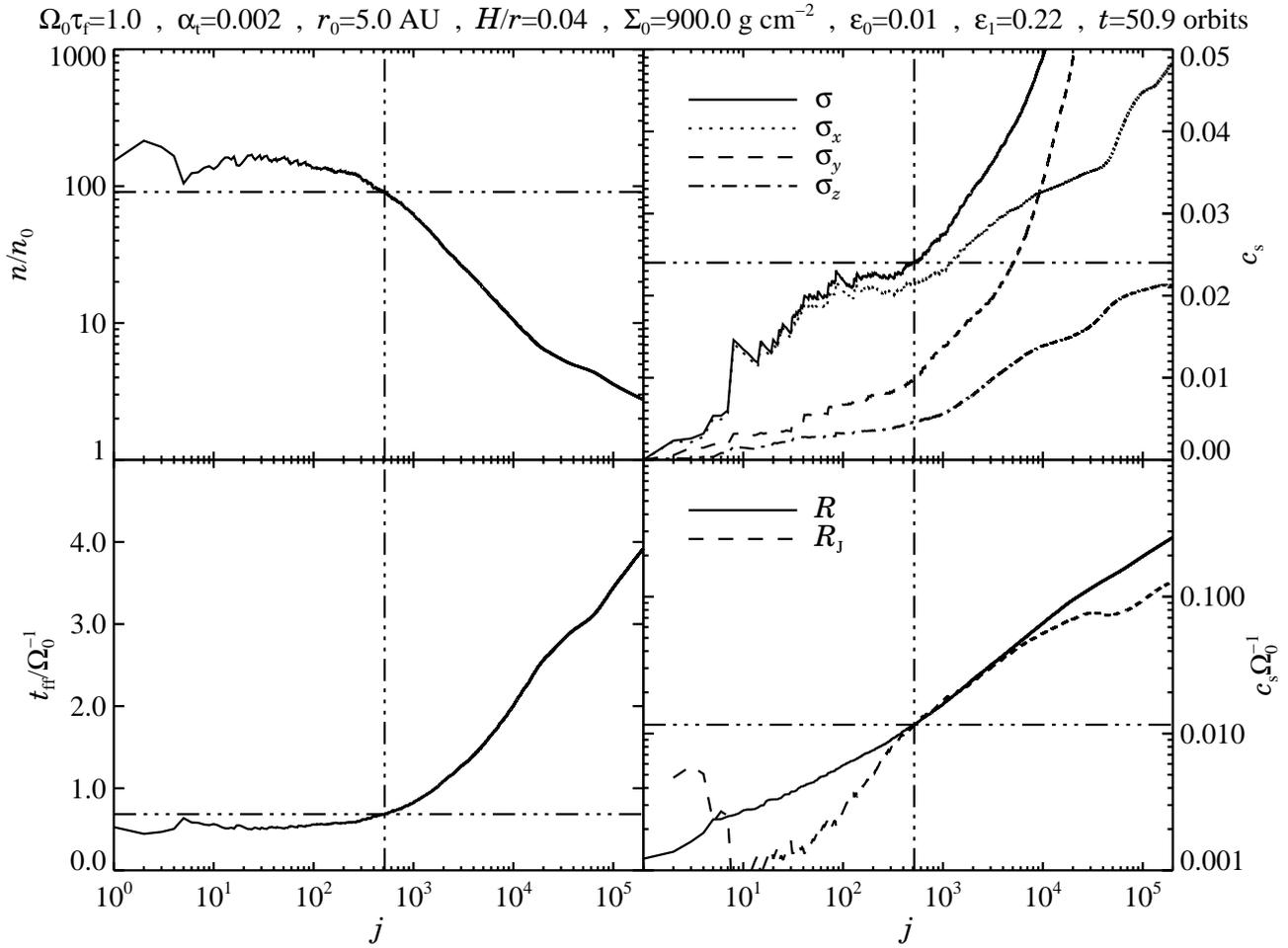}
  \caption{Particle number density $n$ in units of average density $n_0$,
    velocity dispersion $\sigma$ in units of sound speed $c_{\rm s}$, free-fall
    time $t_{\rm ff}$ relative to the clump life-time $t_{\rm cl}$, and clump
    radius $R$ together with Jeans radius $R_{\rm J}$, all as a function of
    the number of included particles around the densest grid point in the box
    at a time of 50.9 orbits of run A. The vertical and horizontal
    dot-dot-dot-dashed lines indicate the regions of gravitational instability
    for the choice of disc model parameters.}
  \label{f:misc_vs_N}
\end{figure*}
\begin{equation}
  \epsilon=\epsilon_1 \exp[-z^2/(2 H_\epsilon^2)] \, ,
  \label{eq:eps_strat}
\end{equation}
with the dust-to-gas ratio scale height given by the expression
$H_\epsilon^2=D_{\rm t}/(\tau_{\rm f} \varOmega_0^2)$.
The dust-to-gas ratio at $z=0$ is
\begin{equation}
  \epsilon_1 = 
      \epsilon_0 \sqrt{ \left( \frac{H}{H_\epsilon} \right)^2 + 1 } \, ,
\end{equation}
where $H=c_{\rm s} \varOmega_0^{-1}$ is the scale height of the gas.
We now proceed by writing the turbulent diffusion coefficient as $D_{\rm t} =
\delta_{\rm t} c_{\rm s}^2 \varOmega_0^{-1}$, where $\delta_{\rm t}$ is the turbulent
diffusion equivalent of $\alpha_{\rm t}$ of \cite{ShakuraSunyaev1973}. Then the
mid-plane dust-to-gas ratio $\epsilon_1$ can be written as
\begin{equation}\label{eq:eps1}
  \frac{\epsilon_1}{\epsilon_0} =
      \sqrt{ \frac{\varOmega_0 \tau_{\rm f}}{\delta_{\rm t}} + 1} \approx
      \sqrt{ \frac{\varOmega_0 \tau_{\rm f}}{\delta_{\rm t}}} \, ,
\end{equation}
where the approximate expression is valid for $\varOmega_0 \tau_{\rm f} \gg
\delta_{\rm t}$. For $\delta_{\rm t} = \alpha_{\rm t} = 0.002$ and
$\varOmega_0 \tau_{\rm f}=1$, this gives $\epsilon_1 \approx 22.4 \epsilon_0$,
so starting from a dust-to-gas ratio of $\epsilon_0=0.01$, the mid-plane
dust-to-gas ratio can be expected to rise to $\epsilon_1=0.22$ due to vertical
settling. Such a low dust-to-gas ratio alone will not for any
physically reasonable
column density cause gravitational
fragmentation \citep{GoldreichWard1973} or be subject to vertical stirring by
the Kelvin-Helmholtz instability \citep[the Richardson number ${\rm Ri}$
is around unity, see e.g.][and stratification with ${\rm Ri}>0.25$
should be stable]{Sekiya1998}. Even at such a high dust-to-gas ratio
we are still in the gas-dominated regime where the back-reaction from
the dust on the gas can be neglected. The turbulent dust concentrations are
assumed to occur in
such a vertically settled dust layer. Now the most overdense regions will
have a dust-to-gas ratio of unity and beyond. But we have measured that only
about 3\% of the grid cells have a dust-to-gas ratio of above unity at any
given time, and thus
it is still reasonable as a first approximation to ignore the back-reaction of
the dust on the gas, although a more advanced study should include this effect
as well.

To find out if a given overdense clump is gravitationally unstable, we shall compare the
different time-scales and length-scales involved in fragmentation by
self-gravity in a Jeans-type stability analysis. First we investigate if the clump is gravitationally bound. We
consider a clump of radius $R$, mass $M$ and velocity dispersion $\sigma$. The
velocity dispersion must include the dispersion due to the background shear. For
such a clump with a given mass to be gravitationally unstable, it must have a
radius that is smaller than the Jeans radius given by
\begin{equation}
  R_{\rm J} = \frac{2 G M}{\sigma^2} \, .
\end{equation}
If this first criterion is fulfilled, then it is also important that the
collapse time-scale of the structure is shorter than the life-time of the
overdense clump $t_{\rm cl}$.
Only then we can be sure that the changing gas flow will not dissolve the
concentration before it has had time to contract significantly.
The fragmentational collapse happens on the
free-fall time-scale
\begin{equation}
  t_{\rm ff} = \sqrt{\frac{R^3}{G M}} \, .
\end{equation}
The condition for gravitational instability is now that $R<R_{\rm J}$ and that
at the same time $t_{\rm ff}<t_{\rm cl}$. We do not have to check
separately that the collapse
happens faster than a shear time $t_{\rm sh}=\varOmega_0^{-1}$, since
the effect of the background shear is already included in the velocity
dispersion.
\begin{figure*}
  \includegraphics{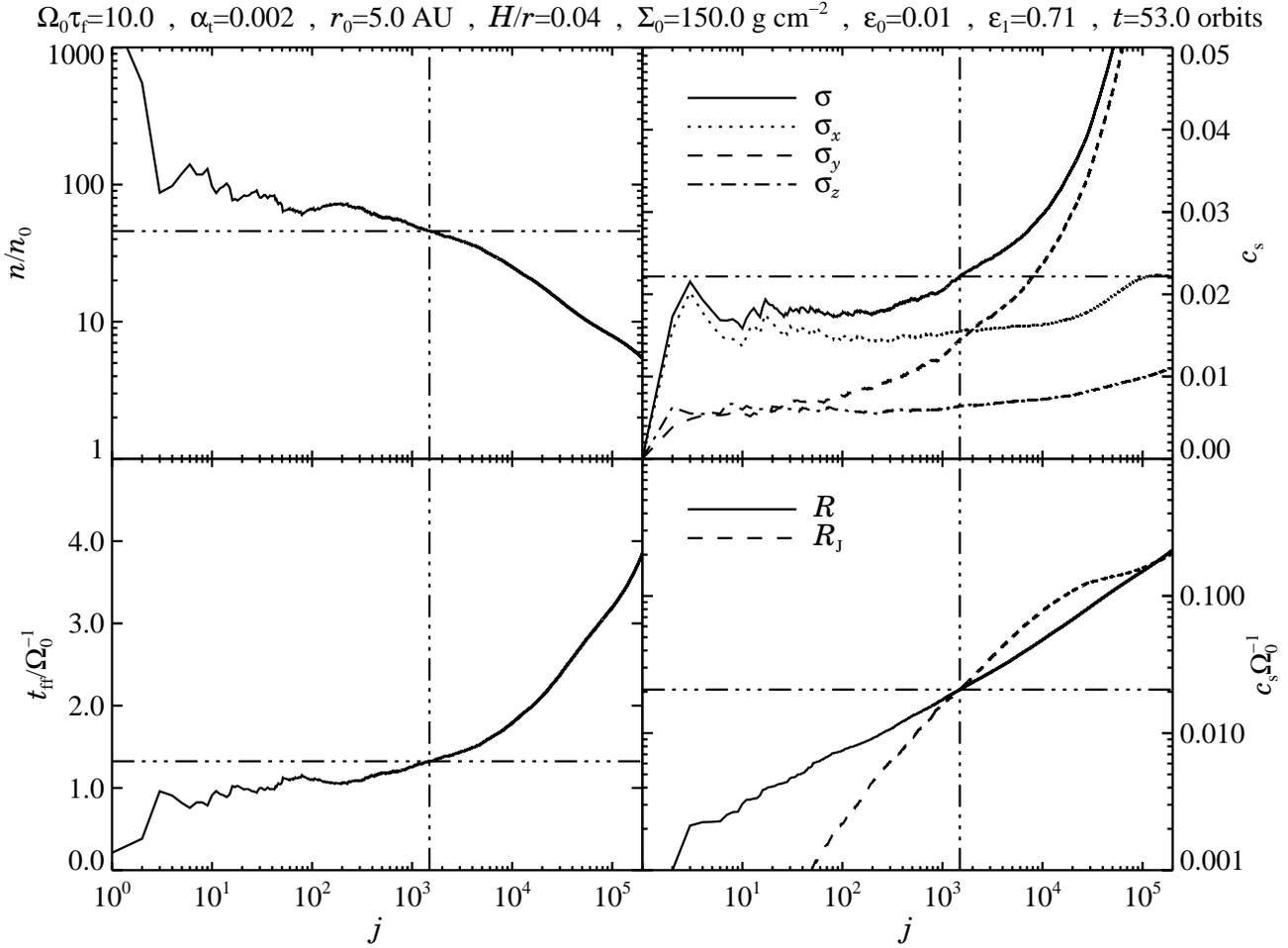}
  \caption{Same as \Fig{f:misc_vs_N}, but for run C (decameter-sized bodies)
    at a time of 53 orbits. Here the minimum mass solar nebula column density
    is sufficient to have a gravitational instability. This is mainly because
    the velocity dispersion is smaller than for run A. Also the high density
    region has a larger extent.}
  \label{f:misc_vs_N_Ot10.0}
\end{figure*}

We now try to find out the 
smallest value of $\varSigma_0$ that gives rise to a gravitational
instability. Then we can see whether this is a value that occurs in nature
or not. For $\varOmega_0 \tau_{\rm f}=1$ (run A) the minimum value of the column
density turns out to be
around $\varSigma_0=900\,{\rm g\,cm}^{-2}$ (6 times the minimum mass
solar nebula value at 5 AU), whereas for $\varOmega_0 \tau_{\rm f}=10$ (run C), a
gravitationally unstable cluster of protoplanetesimals is achieved already at
the minimum mass solar nebula value $\varSigma_0=150\,{\rm g\,cm}^{-2}$.

First run A is considered.  Here we can calculate the mass in each
superparticle. With $\varSigma_0=900 \,{\rm g\,cm}^{-2}$, $\xi=0.04$,
$r=5\,{\rm AU}$ and $\epsilon_1=0.22$, we get $\rho_1=1.2\times10^{-10}\,{\rm
g\,cm}^{-3}$ and $m=8 \times 10^{20}\,{\rm g}$.  Thus each superparticle
represents about $3\times10^{14}$ meter-sized protoplanetesimals. This is five
orders of magnitude more mass than in a kilometer-sized planetesimal, but since
we are interested in identifying gravitationally bound regions with the mass of
thousands and thousands of planetesimals, this is not a problem. Actually
resolving the mass of even one single planetesimal with meter-sized objects
would require on the order of a billion particles, which is way beyond current
computational resources.

We examine the region around the densest grid point of run A
at a time of $t=50.9$ orbits in more detail. This
time is chosen because there occurs a large concentration of particles, see
\Fig{f:npmax_t}. We
consider the $j$ nearest particles to the densest point and calculate for $j$
between $1$ and 200,000 the particle number density $n$, the velocity dispersion
$\sigma$ and its directional components, the free-fall time $t_{\rm ff}$, and
the radius of the clump together with the Jeans radius $R_{\rm J}$. The results are shown in \Fig{f:misc_vs_N}. It is
reasonable to require at least $j=100$ for a measurement to be statistically
significant \citep[for $j\geq100$ the relative counting error falls below 10\%,
see e.g.][]{CasertanoHut1985}. It is also reasonable to require that the size
of the clump be larger than the size of a grid cell, since any
structure in the concentration
within a single grid is not well-resolved. The same is true for the velocity
dispersion.
At $j=100$ the dust number density is more than 130
times the average, but the radius of the $j=100$ clump is only
around 0.007, which is smaller than the grid cell radius of $\delta x/2=0.01$.
At $j\simeq500$ the clump has the size of a grid cell, and here the number
density is more like $100$ times the average.  This must be multiplied by the
enhancement by sedimentation, which is around $20$, to give a dust-to-gas ratio increase by a
factor of $2000$ compared to the original value in the disc. The velocity
dispersion is around $\sigma \sim 0.02\ldots0.03\,c_{\rm s}$. That includes the
velocity dispersion due to the background shear, but this is not a very
important effect anyway because the size of the overdense clump is very small.
At small scales the velocity dispersion is completely dominated by the radial
component, according to \Fig{f:misc_vs_N}, whereas the shear only takes over at larger scales.
\begin{figure*}
  \includegraphics{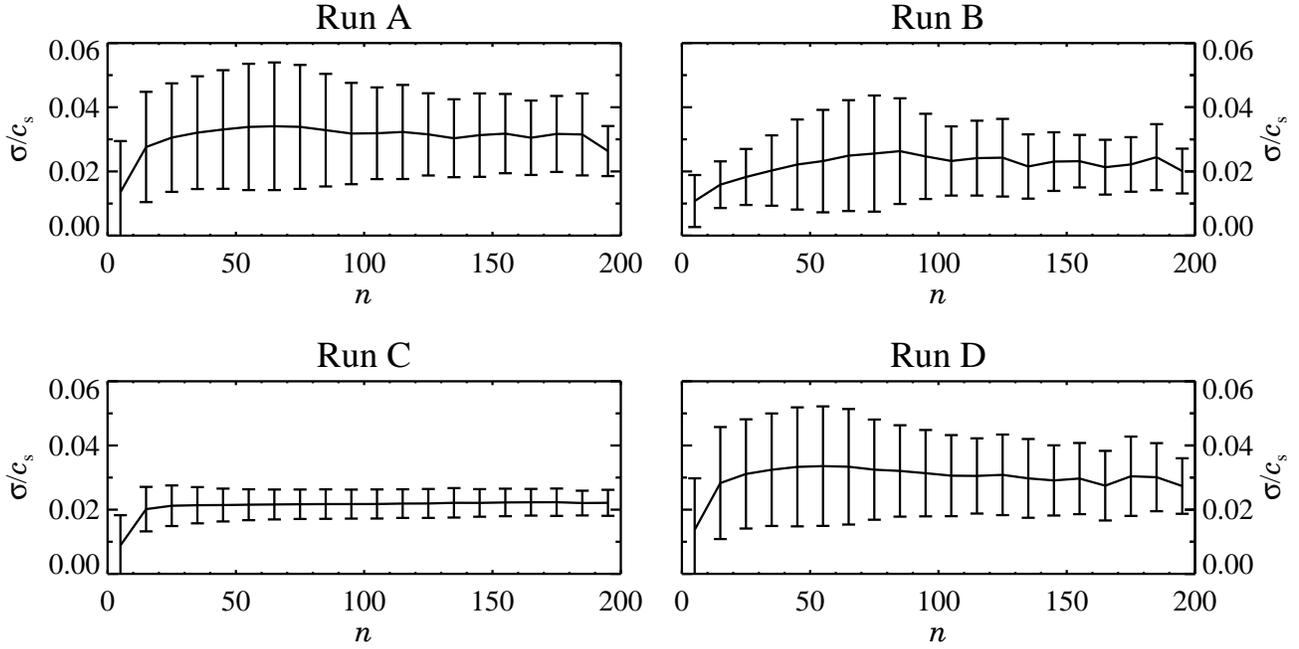}
  \caption{Average velocity dispersion and fluctuation interval as a function
    of the number of particles in a grid cell. The dispersion rises until
    there are around 50 particles in a grid cell, and is then constant up to
    200 particles, or around 30 times the average dust density. This
    corresponds to an isothermal equation of state for the boulders.}
  \label{f:sigmap_n}
\end{figure*}

The free-fall time is a bit below the clump life-time, which is typically one
shear time (see insert in \Fig{f:npmax_t}; note that the time unit is in
orbits). For calculating the Jeans radius we have had to adopt a column density
as high as $\varSigma_0=900\,{\rm g\,cm^{-2}}$ in order to have the clump to be
gravitationally unstable. This is mainly due to the high velocity dispersion.
The radius of the clump is around one Jeans radius at $j=1000$, so the clump is
gravitationally bound at this scale and would be subject to further contraction
by self-gravity. The gravitationally unstable region is around three grid cells
in diameter, but even though this is well within the dissipative scales of the
turbulence, the effect of the unresolved turbulence on the motion of the
particles should be very little, as such small scale turbulence has short
life-times and low amplitudes compared to the large scales. Extrapolating the
resolved large scale turbulence to the grid-scale with a Kolmogorov law gives
lower turbulent velocities than the particle velocity dispersion that we
already measure at the grid scale. Thus we conclude that the unresolved
turbulence has little or no influence on the particle dynamics. The
concentrations and velocity dispersions are exclusively driven by the large
resolved scales of the gas motion.

The solid size of the forming object would be roughly 400 km if all the 1000
superparticles end up in just one large body. On the other hand, the outcome of
such a collapse may also favor the further fragmentation of the clump. This all
depends on how the velocity dispersion behaves with increasing density. In the
$N$-body simulations of \cite{Tanga+etal2004} gas drag works as an efficient
way to dissipate the gravitational energy that is released in the contraction
of protoplanetesimal clusters. Only such simulations, that include self-gravity
and gas drag, could show the further evolution of the overdense boulder clumps
that we see in the present work.

For decameter-sized bodies (run C), we plot in \Fig{f:misc_vs_N_Ot10.0} the same
quantities as in \Fig{f:misc_vs_N} around the densest point at a time of 53
orbits. This time we adopt the minimum mass solar nebula column density of
$\varSigma_0=150\,{\rm g\,cm^{-2}}$, which gives a mid-plane density of $\rho_1=2
\times 10^{-11}\,{\rm g\,cm}^{-3}$.
Because of the high friction time, the
dust-to-gas ratio in the mid-plane (eq.[\ref{eq:eps1}]) is now 0.71. The Richardson number is
correspondingly lower at around ${\rm Ri}=0.4$, so it is still stable to
Kelvin-Helmholtz instability. The mass of the individual superparticles is here
$m=4 \times 10^{20}\,{\rm g}$.
The density is
slightly smaller than for meter-sized bodies, at statistically significant
counts around 50 times the average, but the velocity dispersion is lower,
and also the overdense region is much larger than it was for meter-sized boulders.
Thus already a minimum mass solar nebula can produce a gravitational instability. The
unstable region is as large as 10 grid cells in diameter, and contains
around $10^5$ particles. The size of a solid object consisting of this number
of superparticles is roughly 1,400 kilometers. Again there is also the
possibility that millions of $10$-kilometer objects form instead.

The preceding calculations are of course only an estimation of the potential
importance of self-gravity. In a real protoplanetary disc there will be a
distribution of dust grain sizes present at any time. If e.g. fragmentation is
important, as discussed in the introduction, then the greater part ($80\%$) of
the mass may still be present in bodies that are well below one meter in radius
\citep{DullemondDominik2005}. With only $20\%$ of the mass in the size range
between one and ten meters, the critical column density could be as much as a
factor of two higher than stated above. However, this is still in the range of
the masses derived for circumstellar discs. So the qualitative picture that
the clumps are gravitationally unstable for physically reasonable gas column
densities is robust.

To quantify the velocity dispersion in the entire box, we have calculated the
average values over all the grid cells. The 
results are shown in the last four columns of \Tab{t:simres}. Grid cells with 0
or 1 particles have been excluded from the average because the velocity
dispersion is per definition zero in these underresolved cells. The meter-sized
bodies have the highest velocity dispersion, around
$\sigma_1\approx0.02\,c_{\rm s}$, whereas decimeter bodies have
$\sigma_{0.1}\approx0.014\,c_{\rm s}$ and decameter bodies have a value of
$\sigma_{10}\approx0.017\,c_{\rm s}$. These values are similar to the turbulent
velocities of the gas at the largest scales of the box (see Fig.\ 2 in JK05),
which again shows that these large scales are the drivers of the particle
dynamics.
Interestingly run D, which is similar to
run A only without the radial pressure gradient, has the same velocity
dispersion as run A, so the radial pressure gradient does not add extra
velocity dispersion to the boulders. The toroidal component of the velocity
dispersion is similar for all the runs because it is dominated by the shear
over a grid cell. Run C has a twice as large radial velocity dispersion as run
B. This can be explained because the large particles in run C react much slower
to the local behavior of the gas, and thus particles of different velocities
and histories are mixed in together.

The behavior of the velocity dispersion with increasing dust density is
relevant for gravitational instability calculations. The average velocity dispersion, and
the fluctuation width, as a function of the number of particles in a grid cell
is shown in \Fig{f:sigmap_n}. Again it is evident that the velocity dispersion
for $\Omega_0\tau_{\rm f}$ of unity is largest. For all runs the velocity
dispersion typically rises until there are around 50 particles in the cell.
Then the dispersions stay constant all the way to 200 particles. Thus the
equation of state of the particles is isothermal, at least up to 30 times the
average dust density.

\section{SUMMARY AND DISCUSSION}

We have considered the effect of magnetorotational
turbulence on the motion of dust particles with a freely evolving space
coordinate. The particle treatment was necessary over the fluid treatment,
because the mean free path of the macroscopic dust boulders is so long that they
can no longer be treated as a fluid.
The use of magnetorotational turbulence may not be completely justified in the
mid-plane of the disc where the ionization fraction due to radiation and
cosmic particles is low. But due to its Kolmogorov-like properties, where
energy is injected at the unstable large scales and then cascades down to
smaller and smaller scales, magnetorotational turbulence can be seen as a sort
of ``generic disc turbulence''.

We find that the turbulence acts on the particles by concentrating
meter-sized boulders locally by up to a factor of 100 and by reducing their
radial drift by $40\%$.
Both the concentrations and the reduced radial drift happen because the dust
particles are temporarily trapped in radial density enhancements. One would not
expect such structures to be long-lived in a general turbulent flow, but
magnetorotational turbulence in accretion discs is subject to a strong shear
that favors elongated toroidal structures. In the
presented simulations the typical life-time of the structures is on the order
of a few orbits, corresponding to tens or even hundreds of years in the outer
parts of a protoplanetary disc. When the density structures eventually
dissolve, new structures appear at other locations.
We find a strong correlation between a gas column density of a few percent above
the average and a several times increase in the dust column density. We have
also seen some evidence for increased dust density in regions of
anticyclonicity, but the long friction time of the dust particles makes it
difficult to identify the gas flow that caused a given concentration,
because the concentration may drift away from the creation site.

The large concentrations naturally occur near the grid scale. In finite
resolution computer simulations the dissipative length scale must necessarily
be moved from the extremely small dissipative scales of nature to the smallest
scales of the simulation box. Thus the turbulence is not well-resolved near the
grid scale. On the other hand, the concentrations are driven by the largest
scales of the turbulence, because there are the largest velocities and the
longest lived features \citep{Voelk+etal1980}. Already the other well-resolved
but slightly smaller scales fluctuate too quick and at too low speeds to
influence the path of an object that is one meter in size or larger. This
argument is given support by the fact that we measure particle velocity
dispersions in the grid cells that are comparable to the velocity amplitude of
the gas at the largest scales of the simulation.  Thus, one should not expect
higher resolution to change the concentrations or the velocity dispersions
significantly.

Our estimation of the minimum gas column density that would make the densest
protoplanetesimal clumps gravitationally unstable is necessarily based on many
assumptions. We assumed that half of the dust mass in the disc was present in bodies of
the considered size, whereas in real discs an even larger part of the dust mass may be bound
in small fragments that result from catastrophic collisions. We also ignored
the back-reaction from the dust on the gas.  The background state has, both for
meter and decameter bodies, a dust-to-gas ratio just below unity
(where the back-reaction becomes important). The effect of dust drag on the
magnetorotational instability has to our knowledge never been considered. One
can speculate that the drag force will mimic a strong viscosity and thus
disable the source of turbulence where the dust density is high.  For the
treatment of Kelvin-Helmholtz instability we based it simply on a criterion on
the Richardson number ${\rm Ri}$. There is some indication that this may be too
simplistic and that in protoplanetary discs much higher Richardson numbers are also unstable
\citep{GomezOstriker2005}, but one can also speculate that the full inclusion
of dust particles in simulations of Kelvin-Helmholtz turbulence would show
strong local concentrations like we see here for magnetorotational turbulence.
Thus the exact values of six times the minimum solar nebula for meter-sized
boulders and just the minimum mass solar nebula for decameter-sized boulders
should only be considered as rough estimates.
Still, the result that the clumps are gravitationally unstable for reasonable
gas column densities
is robust enough to warrant further investigations that include treatment of
self-gravity between the boulders.

Thus we find that the gravoturbulent formation of planetesimals from the
fragmentation of an overdense swarm of meter-sized rocks is possible.
Turbulence is in this picture not an obstacle, but rather the ignition spark, as it is
responsible for generating the local gravitationally bound overdensities in the
vertically sedimented layer of boulders.


\acknowledgments

Computer simulations were performed at the Danish Center for Scientific
Computing in Odense and at the RIO cluster at the Rechenzentrum Garching. Our research is partly supported by the European Community's Human
Potential Programme under contract HPRN-CT-2002-00308, PLANETS. We would like
to thank the anonymous referee for a number of useful comments that helped to
greatly improve the original manuscript.

\clearpage

\end{document}